\documentclass{elsart}

\usepackage{epsfig}
\usepackage{amssymb}
\usepackage{pstcol}
\begin{document}
\begin{frontmatter}
\title{New approach of fragment charge correlations 
in $^{129}$Xe+$^{nat}$Sn central collisions
}
\author[Saclay]{J-L.~Charvet}\footnote{Corresponding author. 
{\it E-mail address:} jcharvet@cea.fr (J-L. Charvet),\\
Tel.: 33 169 086068. Fax: 33 169 087584.},
\author[Saclay]{R.~Dayras},
\author[LPC]{D.~Durand},
\author[LPC]{O.~Lopez},
\author[LPC]{D.~Cussol},
\author[Saclay]{D.~Dor\'e},
\author[Saclay]{L.~Nalpas},
\author[LPC]{A.~Van Lauwe},
\author[Saclay]{C.~Volant}

\address[Saclay]{ DAPNIA/SPhN, CEA/Saclay,
91191 Gif-sur-Yvette Cedex, France}
\address[LPC]{LPC Caen,
IN2P3-CNRS/ENSICAEN et Universit\'e,
14050 Caen Cedex , France}

\vskip -0.5cm
\begin{abstract}
A previous analysis of the charge (Z) correlations in the $\Delta Z-<Z>$ plane
for $^{129}$Xe+$^{nat}$Sn central collisions at 32 MeV/n has shown an enhancement in the
production of equally sized fragments (low $\Delta Z$) which was interpreted as 
an evidence for spinodal decomposition. However the signal is weak and rises
the question of the estimation of the uncorrelated yield. 
After a critical analysis of its robustness, we propose in this paper 
a new technique to build the uncorrelated
yield in the charge correlation function. The application of this method to
$^{129}$Xe+$^{nat}$Sn central collision data at
32, 39, 45 and 50 MeV/n does not show any particular enhancement of the 
correlation function in any $\Delta Z$ bin.

\end{abstract}

\end{frontmatter}

\section{Introduction}

The search for signals of a liquid-gas phase transition in the nuclear medium is presently
a very active field of research (see \cite{chomaz1} and ref. therein). 
In such a framework, multifragmentation is a key process. A possible scenario is
the excursion of the excited
system, produced in dissipative heavy-ion collisions, 
at low densities inside the coexistence region of the nuclear phase diagram 
(for a general introduction to the subject, see for instance \cite{durand}).
There, a volume instability, the so-called {\it spinodal
decomposition}, leads to a disassembly of the system \cite{bertsch,ayik}.
Such a behaviour is predicted by microscopic transport calculations
based on the nuclear Boltzmann equation and its stochastic extensions \cite{chomaz2,bertrand,guarnera}. 
The main idea is to search for fragmentation
events in which most fragment atomic numbers are nearly
equal as suggested by theoretical works \cite{chomaz3,guarnera96}.
It is worth noting that surface instabilities, like Rayleigh or sheet instabilities \cite{moretto2,bauer},
may produce, in the early compression-expansion stage of the collision, 
the formation of toroidal and bubble nuclei (depending of the
value of the incompressibility constant $K$) that may also decay into several nearly equal fragments \cite{xu}. In any case,
the proportion of this kind of multifragmentation events should be small and a sensitive method, called ``higher-order correlations'',
has been proposed in Ref. \cite{moretto} to search for weak signals in the fragment charge correlation analyses.
Following the prescriptions of this method, the question of a possible experimental signature of spinodal decomposition 
of excited nuclear matter has been examined for $^{129}$Xe+$^{nat}$Sn central collisions at 32 MeV/n 
in a previous work \cite{borderie}.

The analysis relies on two variables. 
The average fragment charge per event is defined
as:
\begin{equation}\label{eq1}
<Z>=\frac {1}{M} \sum_{i=1}^{M} Z_i 
\end{equation}
where $Z_i$ is the charge of fragment $i$ and $M$ is the fragment multiplicity. 
Only fragments with atomic number $Z_i \geq 5$ are considered. As in
Refs \cite{moretto,borderie}, the standard deviation in fragment sizes is defined
as:
\begin{equation}\label{eq2}
\Delta Z=\sqrt {\frac {1}{M-1} \sum_{i=1}^{M} (Z_i-<Z>)^{2}}
\end{equation}
The correlation function, $R$, is built by considering the ratio of the measured
yield $Y(\Delta Z, <Z>)$ versus the ``uncorrelated'' yield $Y'(\Delta Z, <Z>)$. Since events with 
nearly equal size fragments are associated with low values of $\Delta Z$, any significant
enhancement of $R$ for small $\Delta Z$ ($\Delta Z<1$) may signal a spinodal decomposition
as proposed in \cite{borderie}.
The key question is how to build the uncorrelated yield $Y'$.
This last quantity has been estimated 
for each class of fragment multiplicities by mixing fragments from different events of the selected sample \cite{moretto,borderie}.
In doing so, the initial total charge carried out by the fragments ($Z_{bound}=\sum_{i=1}^{M} Z_i=M <Z>$)
and also the average fragment charge $<Z>$
are not conserved in building $Y'$. 
Consequently, in the absence of correlations
the ratio $R=Y/Y'$ is not equal to unity and depends simultaneously on the $<Z>$ and $\Delta Z$ values.
Thus, to extract the true correlation one has to determine a background function $R'$ 
corresponding to the ratio $Y/Y'$ for uncorrelated events. 

In the present paper, in a first step, we would like to show how the method
previously used to generate the charge correlation functions \cite{borderie}, that we will call $Z_{bound}$ mixing method, 
may generate spurious signals. 
Then, a new method to build the uncorrelated yield $Y'$ is presented in Section 3. By applying this method, 
the fragment charge correlations obtained in the $^{129}$Xe+$^{nat}$Sn central collisions between 32 and 50 MeV/n are discussed in Section 4.
A recently proposed method to build the denominator $Y'$, called IPM for Intrinsic Probability Method \cite{pierre}, 
which has been applied to the same Xe+Sn data set \cite{tabacaru}, is commented in Section 5. 

\section{Review of the $Z_{bound}$ mixing method}

\subsection{Data selection and previous charge correlation results}

 The charge correlations were analysed in a set of events
 associated with a ``quasi-fusion'' process occurring in central $^{129}$Xe+$^{nat}$Sn collisions at 32~MeV/n.
 Only events with the sum $Z_{tot}$ of the detected charges greater or equal to
 80 were considered,
 the total charge of the Xe+Sn system being 104. Briefly, the event selection was based on the measurement of the
 distribution of the preferred emission direction of matter in the centre of mass of the
 reaction (flow angle). In this work, fragment charge correlation analyses 
 have been  performed by selecting events having a flow angle larger than
 $60^{o}$. It has been shown \cite{marie,frankland} that such events
 ($\Theta_{flow} \geq 60^{o}$) could be interpreted 
 as resulting from the multifragmentation of a thermalised and compact nuclear system
 (``single source'' events). 
 In a previous work \cite{borderie}, using the same gating condition, 
 the charge correlation functions $R(\Delta
 Z, <Z>)$ were first
 built for each fragment multiplicity $M=$ 3, 4, 5 and 6.
 However, the results suffered from a lack of statistics.
 The numbers of events in the first bin in $\Delta Z$ ($\Delta Z < 1$)
 and for each multiplicity of this selection are shown in the third column of Table 1.
 As we see, these numbers, summed over all $<Z>$, are quite small, 
 specially for multiplicities $M=$ 5 and 6. Thus, to increase the statistical significance of the signal,  
  the charge correlations were built 
 by replacing the variable $<Z>$ by the sum of the fragment charges $Z_{bound}$ 
 (each bin in $Z_{bound}$ containing six atomic numbers). Summing over all
 $M$, the resulting
 correlation functions shown in
  Fig. 3 of Ref. \cite{borderie} were interpreted 
 as an evidence of a significant enhancement of equal-sized fragments associated with fused events in Xe+Sn collisions at 32 MeV/n.
 However, as we mentioned previously, the charge correlation function cannot be
 compared to unity but to a specific $R'$ background function.
 This function has been estimated in Ref. \cite{borderie} by assuming that, 
 in the absence of correlations and for a given $Z_{bound}$, $R'=f(\Delta Z,Z_{bound})$ is exponentially increasing with $\Delta Z$ 
 in the range $0<\Delta Z<8$. Obviously, the choice of the background estimation is the critical point of this
 method and in order to test the hypothesis of Ref. \cite{borderie} we have performed a charge correlation analysis $R(\Delta Z, Z_{bound})$
 from a sample of randomly generated events. 

\subsection{Random break-up simulation}

 In the Monte Carlo simulation, the following procedure, that we will call random break-up, has been used:
 for a given fragment multiplicity $M$, the charge of the fragments is obtained by breaking $M-1$ bonds at random in a
 chain of length $Z_{bound}$. Only events for which all fragments have $Z \geq 5$ were kept.
 For each multiplicity, the $Z_{bound}$ distribution was the experimental one for central
 Xe+Sn collisions at 32 MeV/n. We have checked that for each multiplicity $M$, the ensemble thus generated
 is equivalent to the one obtained by computing all partitions and their permutations containing $M$ fragments
 with $Z_i \geq 5$ such that $\sum_{i=1}^{M} Z_i=Z_{bound}$ \cite{dayras}. The $Z_{bound}$ distributions are given
 in the first column of Fig. \ref{fig1} as a function of the fragment multiplicity $M$. They are gaussian in shape
 with a centroid which increases from $\simeq$54 for $M$=3 to $\simeq$64 for $M$=6, whereas their width is practically
 constant. To minimize statistical effects, twenty times more events were generated in the simulation than in the data.
 The simulated $Z$-distributions (full lines) are compared to the experimental data in the second column of Fig. \ref{fig1}.
 For multiplicities $M$= 5 and 6, the agreement is remarkably good. At lower multiplicities $M$= 3 and 4, the simulated 
 distributions are slightly broader than the experimental ones. 
 The last
 two columns of Fig. \ref{fig1} compare the simulated $\Delta Z$
 distributions (full lines) to the experimental ones with two $\Delta Z$ scales
 (in order to show the details at low $\Delta Z$). 
 There again for multiplicities 5 and 6, the agreement between the
 simulation and the data is remarkable. For the lower multiplicities 3 and 4, as for the $Z$-distributions, the
 simulated distributions are broader than the experimental ones. 
 Examination of the last column of  Fig. \ref{fig1} reveals no 
 enhancement over the simulation for $\Delta Z < 6$.
 The numbers of events with $\Delta
 Z<1$  obtained in the simulation are listed in the second column of Table 1 as a function of
 multiplicity. We note that the total fraction of events (0.154 $\pm$ 0.005 \%) in this $\Delta Z$
 bin is compatible with the experimental one (0.127 $\pm$ 0.020 \%).
 
 Fig. \ref{fig2}a shows the ratio between the random
 break-up yield $Y_{break-up}$ over $Y_{partitions}$ obtained by calculating analytically all possible partitions
 \cite{dayras} as a function of 
 $\Delta Z$ for each multiplicity $M$
 and their sum. 
 Except for the low statistics data with large $\Delta Z$ ($M$=6),
all of the ratio values are consistent with unity.
It is worth noting that in the present case, deviations from unity are of statistical origin only.
Fig.\ref{fig2}b displays the values of $Y_{break-up}/Y_{partitions}$ in the bin
 $\Delta Z<1$ as a fonction of $<Z>$. 
 In this bin, the statistics is poor  
 and we observe deviations in excess of one standard deviation ($\sigma$).
 In the following, the random break-up events are analysed using 
 the $Z_{bound}$ mixing method \cite{borderie}. 
 
\subsection{Application of the $Z_{bound}$ mixing method to the random break-up
simulation}

 Firstly, the charge correlation functions $R(\Delta Z,<Z>)$ were built
 for each multiplicity $M=3,4,5,6$, separately. For this purpose, the uncorrelated yield $Y'$ was built 
 generating $10^3$ times more events than in the numerator in order to minimize statistical 
 errors on the denominator.
 The $R$ values corresponding to four ($M,<Z>$) pairs chosen in the interval $48 \leq Z_{bound} \leq 52$
 are shown in Fig. \ref{fig3}a. They illustrate the general evolution of the function $R=Y/Y'$.
 The amplitude and extension in $\Delta Z$ of the function $R$ depend on the
 multiplicity $M$ and the average charge $<Z>$. 
 However, we note that within statistics, the charge correlation functions
 decrease with decreasing $\Delta Z$ without any significant enhancement in the first bin in $\Delta Z$,
 as expected for uncorrelated events. 
 In the absence of correlations other than charge conservation,
 these results (Fig. \ref{fig3}a) represent the background function $R'$ discussed above.
 Large fluctuations and statistical error bars are present in the low $\Delta Z$ bins. 
 Thus, a correct estimate of the function $R'$ from experimental data, often associated with weaker 
 statistics, appears to be a very difficult task.
 
 Secondly, following \cite{borderie}, we built the charge correlations $R(\Delta Z, Z_{bound})$.
 They are displayed in Fig. \ref{fig3}b for six values of
 $Z_{bound}=39,45,51,57,63,69$ as a function of $\Delta Z$ (each bin in $Z_{bound}$ regrouping six atomic numbers).
 The four ($M,<Z>$) pairs shown in Fig. \ref{fig3}a are now integrated in the bin $Z_{bound}=51$.
 Obviously, our events being uncorrelated all these correlation functions are
 the background functions $R'$. Contrary to what was
 assumed in \cite{borderie} these background functions are not exponential
 functions of $\Delta Z$ even in the restricted range $\Delta Z<8$. 
 Thus, if as in \cite{borderie}, one assumes an exponential background in the range $2<\Delta
 Z<8$ (the lines in Fig. \ref{fig3}b), and extrapolates it down to zero, one
 creates artificially a signal at low $\Delta Z$.
 The fragment charge correlations for the bin $\Delta Z<1$ and the corresponding background extracted this way (broken
 line) are plotted in Fig. \ref{fig4} (up-left)
 as a function of $Z_{bound}$.
 This figure is very similar to Fig. 3 of Ref. \cite{borderie}.
 This spurious result finds its origin in the mixing of events with different
 fragment multiplicities $M=$ 3, 4, 5 and 6 to build $Z_{bound}$ bins.
 Indeed, the charge correlation functions $R(\Delta Z, Z_{bound})$ combine various functions $R(\Delta Z, <Z>)$ which 
 are multiplicity dependent. 
 
 This artificial enhancement of equal-sized fragments is not 
 systematic and depends on the $Z_{bound}$ distributions. 
 Xe+Sn central collisions have also been studied with INDRA
 at 39, 45 and 50 MeV/n \cite{salou,borderie2}. In Ref. \cite{borderie2}, the authors show
 the disappearence of the charge correlation signal at low $\Delta Z$ when the energy increases. But, the same 
 evolution is obtained by analysing simulated data without charge correlations.
 By using the $Z_{bound}$ mixing method, we show in Fig. \ref{fig4} the charge correlations in the first $\Delta Z$ 
 bin of four sets of events generated by the
 random break-up simulations 
 obtained from the $Z_{bound}$ distributions of Xe+Sn data at 32, 39, 45 and 50 MeV/n (with 20 times the statistics of the data). 
 This figure, very similar to Fig.2 of Ref. \cite{borderie2}, 
 shows clearly the artificial decreasing of the charge correlation signal when the energy increases. 
 Consequently, the method using the mixing of multiplicities \cite{borderie} renders unreliable an estimation of the background
 $R'$.

\section{The exchange mixing event method}

\subsection{Description of the method}

To get rid of the problematic evaluation of the background, 
we attempted to build an uncorrelated yield $Y'$ with the conservation of 
the total fragment charge ($Z_{bound}$). In such a case the charge correlation function can be directly compared to unity.
The following procedure has been developed.
For an ensemble of events with a given fragment multiplicity, $M$, we consider
the exchange of two fragments of charges $Z_1(i)$ and $Z_2(i)$ belonging to the event labelled $i$ 
with two fragments of charges $Z'_1(j)$ and $Z'_2(j)$ belonging to the event labelled $j$ under the condition
that $Z_1(i)+Z_2(i)=Z'_1(j)+Z'_2(j)$. As such, the total fragment charge $Z_{bound}$ and the 
average charge $<Z>$ are conserved 
event by event and then the $Z_{bound}$ and the
charge distributions of the considered ensemble are conserved. 
For a given event $i$, we performed the substitutions
by a random picking of event $j$ followed by a random picking of charges inside, respectively, the events $i$ and $j$.
If the condition $Z_1(i)+Z_2(i)=Z'_1(j)+Z'_2(j)$ was verified the charge substitutions were performed, if not 
we proceeded with a new random picking of event $j$, and so on,
until all events in the sample has been changed at least once (one iteration).
First, 100 iterations are performed to get a mixed sample. 
Then, the yield $Y'(\Delta Z, <Z>)$ is created from this mixed sample by continuing the same iterative procedure $10^3$ times  
and at each time $Y'$ is incremented in order to make negligible its statistical error (relative to $Y$).
Accurate tests showing the sensitivity of the exchange method 
are developed in the next.
One can also consider exchanges among the events in which more than two (say $N$) 
fragments are exchanged. In this latter case, the method is obviously restricted to events for 
which $M$ is larger than $N$ and in the following we only use $N=2$.
 
\subsection{Tests of the exchange mixing method}

The technique has been tested with the random break-up 
calculation and the results are plotted in Fig. \ref{fig5}a for each multiplicity and the whole ensemble
$M=3,4,5,6$. For any multiplicity, the
correlation function $R=f(\Delta Z)$ is constant and close to unity 
(most of the statistical error bars are smaller than the symbol size). 
Although we performed this simulation 
with a statistics twenty times larger than the one of the Xe+Sn data at 32 MeV/n (see Table 1), 
statistical fluctuations are still present, specially, in the first and largest bins in $\Delta Z$ for $M=$ 6.
The correlation function $R$ corresponding to the whole multiplicity set is equal to unity 
(bottom picture in Fig. \ref{fig5}a) independently
of $\Delta Z$. In this latter case, the fact that the correlations $R=f(\Delta Z)$ are 
multiplicity independent validates the multiplicity regrouping.

However, when this method is applied to the data,
the denominator may partially keep the fragment charge correlations present 
in the experimental sample (numerator). For example, applied on an event sample populated in majority by nearly-equal
sized fragments, this method, as any others based on the data, would paradoxally give a weak signal. So, the exchange method may
only be relevant in the search of a weak proportion of equal-sized fragment events in the data and it is
crucial to test its sensitivity to measure    
an enhancement of charge correlations precisely in the bin $\Delta Z<1$.
In order to do this test, we doubled the number of events populating the first bin in $\Delta Z$ in 
the random break-up sample and build new uncorrelated yields $Y'$.
The results of the charge correlation analyses of this new sample are displayed in Fig. \ref{fig5}b.
As expected, the charge correlations $R$
in the first bin, within statistical uncertainties, are twice the ones shown in Fig. \ref{fig5}a, whereas they are unchanged in other
bins. By integrating over all multiplicities $M$ and $<Z>$, the correlation value in the first bin is: $R=1.941 \pm 0.045$ (bottom picture in
Fig. \ref{fig5}b). In this case, the number of events is large and we performed the same analysis with a set of simulated events
statistically compatible with the Xe+Sn data at 32 MeV/u.  In the first bin, the new correlation value
$R=1.961 \pm 0.192$ is very close to the expected value $R=2$ at $\simeq5$ standard deviations from one.
These tests substantiate the good sensitivity of the exchange procedure used to build $Y'$.

The use of the random break-up simulation is also a powerful tool to evaluate the kind of errors induced by 
the exchange mixing method. In such a simulation, where no correlation is expected, the deviations from unity
should contain a statistical part due to the number of generated events (numerator), called $\sigma_{N}$,
and, perhaps, a systematic error due to the method itself. 
In Fig. \ref{fig6}a,  we show the deviations from 1 of all charge correlation
values $R(\Delta Z, <Z>, M)$, normalized to their statistical error bar $\sigma_{N}$ (absissa $x=(R-1)/\sigma_{N}$), obtained from 
the simulations previously described and for each multiplicity $M=3,4,5$ and 6 (see Fig. \ref{fig5}a).
The good superimposition of the $x$ distribution with the normal function (see Fig. \ref{fig6}a) shows clearly
that the errors associated with the exchange method are purely statistical.
As all possible partitions ($Y_{partitions}$) have been analytically calculated (see Fig. \ref{fig2}) 
in the random break-up simulation, the yield $Y'=Y_{partitions}$ has been used as denominator in Fig. \ref{fig6}b.
The remarkable similarity between both $x$ distributions in Fig. \ref{fig6}a and \ref{fig6}b indicates that there is no 
contribution due to the exchange method itself in the fluctuations of $R$.
In the same manner, we plotted in Fig. \ref{fig6}c, 
the distribution of $x$ obtained from the Xe+Sn data at 32 MeV/n. In this data, the good superimposition 
of the distribution of $R(\Delta Z, <Z>, M)$ values by the normal function,
indicates that all deviations of $R$ from unity appear compatible with statistical fluctuations. 
In the next Section, the fragment charge correlation functions associated to the Xe+Sn data are  
presented as a function of $\Delta Z$, $<Z>$ and $M$. 

\section{Application of the exchange mixing method to the Xe+Sn data}

In the Xe+Sn central collisions, the multifragmentation process is present at 32, 39, 45 and 50 MeV/n 
and complete events ($Z_{tot} \geq 80$) selected with $\Theta_{flow} \geq 60^{o}$
have been considered to arise from the multifragmentation of single nuclear sources \cite{marie,salou}.
For a given fragment multiplicity, the number of events with $\Delta Z<1$ at respectively 39, 45
and 50 MeV/n bombarding energies are listed in columns 4 to 6 of Table 1. The production of near equal-sized 
fragment events increases with the incident energy and reaches $\simeq2$\% at 50~MeV/n. 
This trend is a result of charge conservation. The multiplicity of light charged particles and fragments ($Z\leq4$) 
is increasing with the incident energy \cite{metivier}. Hence, 
the average total fragment charge ($Z_{bound}$) decreases. Thus for a given fragment multiplicity,
the size of the fragments decreases and the probability to measure
equal-sized fragment events naturally increases.
The analyses of the charge correlations for these ``single source''
samples have been performed, using the exchange mixing procedure to generate $Y'$. 
The charge correlations $R$ have been built at each energy.
First, we show in Fig. \ref{fig7} the charge correlation functions $R=Y/Y'$, summed over all $<Z>$, as a
function of $\Delta Z$  for each multiplicity $M=$ 3, 4, 5 and 6 (the first four rows) and for each incident energy (columns).
The bottom row of Fig. \ref{fig7} 
corresponds, at each bombarding energy, to the whole experimental data (summed
over $M$) 
and benefits from enhanced statistics. All values of $R$ are close to unity.
To be more quantitative, Fig. \ref{fig8} displays 
the quantity $(R-1)/\sigma$ as a function of $\Delta Z$
for random break-up (with analytical background) simulated events (first column) and 
for the experimental data from 32 to 50~MeV/n (columns 2 to 5), for 
multiplicities $M=3$ to 6 (rows 1 to 4) and for all multiplicities (last row).
For convenience, we have drawn dotted lines corresponding, respectively, to one and two standard
deviations. We note that the fluctuations around zero are about of the same amplitude in the simulation and in the data over
the whole range in $\Delta Z$. The fluctuations for small $\Delta Z$ (as well as for large $\Delta Z$) are not particularly larger than in any
other $\Delta Z$ bins.
Nevertheless, following theoretical predictions \cite{chomaz3,guarnera96} the overproduction of 
equal-sized fragments should specifically concern a narrow range in $Z \simeq 10-15$.
If true, the summing over all $<Z>$ to determine the correlation functions may reduce significantly the importance of the ``signal''.
For this reason, we show in Fig. \ref{fig9} the $R$ values corresponding to this first bin as a function of $<Z>$
for each multiplicity $M=$ 3, 4, 5 and 6 (the first four rows), for all multiplicities (last row)
and for each incident energy (columns).
All $<Z>$ bins populated with more than one event have been plotted in Fig. \ref{fig9}. This choice
induces the absence of data point at 32 MeV/n for $M=$5 and 6. Notice that experimental error bars are statistical only.
All values of $R$ are within $2\sigma$ of 1.
It has to be emphasized that the $R$ values (Fig. \ref{fig9}) for each
multiplicity corresponding to the energies 45 and 50 MeV/n are 
particularly close to the $R=1$ line and associated with small statistical fluctuations.
On the other hand, at 32 MeV/n and 39 MeV/n, we are dealing with large statistical errors.
The, perhaps, systematic enhancements of $R$ at 39 MeV/n for multiplicity 4 in the range $<Z>=7-11$,
are compatible with 1 within $1.2\sigma$ which is poorly significant (less than 80\% of confidence level). 
Finally, we do not observe 
any significant enhancement of equal-sized fragments in the Xe+Sn central collisions between 32 and 50 MeV/n.
 
\section{Comparison between methods}

Recently, an alternative method, called IPM \cite{pierre}, 
has been proposed to build the denominator $Y'$ using the independent emission hypothesis
constrained by the charge conservation. 
Basically, it is assumed that the probability to observe a given partition
($n_{z}: (n_{1},..., n_{Z_{tot}})$) with a multiplicity $M$ of a nucleus of charge $Z_{tot}$ is given by the multinomial formula 
\begin{equation}\label{eq3}
P(\{n_{z}\})= \alpha M! \prod \frac {(P_{z})^{n_{z}}}{n_{z}!} \mbox{     with     } {\sum_{z} {P_{z}}=1}
\end{equation}
where $\alpha$ is the normalization constant ($\sum_{n_{z}} {P(\{n_{z}\})}=1$),
$P_{z}$ is the 
intrinsic probability for emitting the charge $z$, $n_{z}$ the number of charges $z$ in 
the partition ($Z_{tot}=\sum_{z} {n_{z}}{z}$) and $M$ the multiplicity of the partition ($M=\sum_{z} {n_{z}}$). 
We note that in the case of a fully random break-up, the partitions are weighted by the factor $M!/ \prod {n_{z}!}$.
Thus, in that particular case the quantity $\prod (P_{z})^{n_{z}}$ is a constant.
For large $Z_{tot}$, the normalization condition implies $P_{z} = 2^{-z}$ \cite{pierre}. 
Any other laws for the $P_{z}$ will imply a departure from random break-up and 
the partitions thus generated might not be free from correlations.
In the IPM method, the probabilities $P_{z}$ are determined by fitting the partition probabilities $P(\{n_{z}\})$ of the numerator
and, as a consequence, they may keep memory of the experimental correlations.
In other words, the IPM method may measure charge correlation deviations from 
the independent emission  hypothesis but the absence of charge correlations
in the denominator $Y'$ is not obvious.
{Moreover, as the IPM method relies upon a fit to the statistical weights associated with each fragment
partition, it washes partly out statistical fluctuations. In Ref. \cite{tabacaru}, the IPM method has been applied to the Xe+Sn data and
the deviations of $R$ from unity 
are distributed according to a law which is narrower than a normal distribution. However what should be this law is
unknown. Thus, the fact that the fluctuations of $R$ around unity are narrower in the case of the IPM method compared with 
the exchange mixing method
does not guarantee that the former method is more sensitive.
The question of sensitivity has been accurately discussed for the exchange mixing method 
by performing several tests from simulated event samples in which the amplitudes of charge correlations 
in the bin $\Delta Z<1$ were precisely known (see Section 3.2).

In \cite{pierre,tabacaru}, the IPM method has been applied to a set of partitions generated by the
Brownian One-Body (BOB) dynamical model
which supports the spinodal decomposition of hot and dilute finite nuclear systems
\cite{chomaz2,guarnera}. This calculation, performed for head-on Xe+Sn 
collisions at 32 MeV/n, has been shown to reproduce quite well the experimental fragment multiplicity
and charge distributions \cite{rivet,frankland2}. 
Of the 1\% of events in the bin $\Delta Z<1$, a percentage of 0.36\% of events
has been evaluated in excess of the background with the IPM method
which represents a correlation value $R\simeq1.5$ \cite{tabacaru}.
The charge correlation function of the same BOB event sample  
has been also determined by using the exchange mixing method proposed in this work and, 
inside the statistical error bars, 
no enhancement of equal-sized fragments was visible \cite{rivet2}. This contradictory result is really puzzling.
Indeed, from the simulated event sample obtained from the $Z_{bound}$ distribution of the Xe+Sn data at 32 MeV/n
(see Section 3.2), with a statistics comparable to the BOB calculation one, 
we increased the number of events in the first $\Delta Z$ bin by 50\%. 
We analysed, by taking the  
definition $\Delta Z=\sqrt {\frac {1}{M} \sum_{i} (Z_i-<Z>)^{2}}$ used in Ref. \cite{tabacaru}, the fragment charge correlations 
of this new set of events with the exchange method. In the first $\Delta Z$ bin the
charge correlation value, integrated over all $M$ and $<Z>$, was: $R=1.450\pm 0.121$, which is compatible with the expected value
$R=1.5$ at $3.7\sigma$ from $R=1$.
Once more, we conclude about the good ability of the exchange method to exhibit a correlation signal with a low rate of overproduction 
of equal-sized fragment events. In fact, in the BOB calculation
equal-sized fragment events are produced early in the collision when the system goes through the
spinodal region of the phase diagram of the nuclear matter. Then, subsequent steps like coalescence to form heavier fragments,
thermal desexcitation and detection filtering, coupled with the finite size of the system, 
deeply blur the original picture of the partitions \cite{frankland2}.
Thus, it seems difficult to use BOB calculation as a reference to assess the sensitivity
of the various correlation methods.

We note that, applied to the same Xe+Sn data analysed in this work, the IPM method
shows an enhancement of equal-sized fragment production at 32, 39 and 45 MeV/n \cite{tabacaru}, in contradiction with
our  results.

\section{Conclusions}

We have developed a two-step analysis to address the issue of charge correlations in nuclear collisions. 
First, we have shown, by using a simulation which randomly generates fragments, 
how the extrapolation to low values of the $\Delta Z$ distribution as
performed in \cite{borderie} may induce a spurious enhancement of the correlation function at small $\Delta Z$.
We have underlined the difficulties to make a
quantitative analysis of the charge
correlations when the total fragment charge ($Z_{bound}$) is not conserved in building the uncorrelated yield $Y'$.
Then, to avoid such a problem, we have discussed
a new method to build the yield $Y'$ by mixing events under the constraints of the conservation of the
total charge of the fragments and the invariance of charge distributions for a given fragment multiplicity.
This method has been successfully tested with a random break-up simulation and is sensitive to any statistically significant 
deviation from a pure uncorrelated fragmentation process. 
Finally, it has been applied in multifragmentation samples of the $^{129}$Xe+$^{nat}$Sn central collisions at 
energies between 32 and 50 MeV/n. All charge correlation functions at low $\Delta Z$ are equal to unity
within two standard deviations and we conclude that no significant enhancement of equal-sized 
fragments has been observed.
These results are in agreement with those obtained in the reactions Xe+Cu and Ar+Au at
50 MeV/n \cite{moretto} but at variance with other analyses of the Xe+Sn data 
using different estimations of the uncorrelated yield $Y'$ \cite{borderie,tabacaru}.
In view of these contradictory results, caution should be taken before giving any definite conclusion on
a signature of spinodal decomposition in nuclear systems by charge correlation measurements.

The experiments $^{129}$Xe+$^{nat}$Sn were performed at GANIL by the INDRA collaboration.

\newpage 

\begin{table} 
\caption{Number of events in the bin $\Delta Z<1$ 
for each multiplicity and their sum for the random break-up calculation and
the Xe+Sn ``single source'' selection at four energies.
}
\vspace*{.2cm}
\begin{center}
\begin{tabular}{l l l l l l}
\hline
 { }  &  {Break-up} &  {32 MeV/n} &  {39 MeV/n}
&  {45 MeV/n} &  {50 MeV/n} \\
\hline
M=3      & 506 & 20    & 22   &  45    & 85  \\	  
M=4      & 382 & 16    & 46   &  72    & 194 \\	  
M=5      & 100 &  5    & 15   &  53    & 145 \\	  
M=6      & 25 &  1    & 11   &  39    & 83    \\	  
\hline
Sum      & 1013  & 42    & 94   &  209    & 507 \\	  
\hline
NT       & 659260 & 32963    & 28088   &  25870    & 26033 \\	  
\hline
Sum/NT        & 0.154\% & 0.127\%    & 0.335\%    &  0.808\%    & 1.948\% \\	  
\hline
\end{tabular}
\end{center}
\vskip 0.5cm
\noindent
The two last rows give respectively the total number of events (NT) and the percentage of 
events in the bin $\Delta Z<1$. The break-up calculation with 20 times the
statistics has to be compared with the Xe+Sn data at 32 MeV/n.
\end{table}
 
\newpage

\begin{figure} 
\begin{center}
\psfig{file=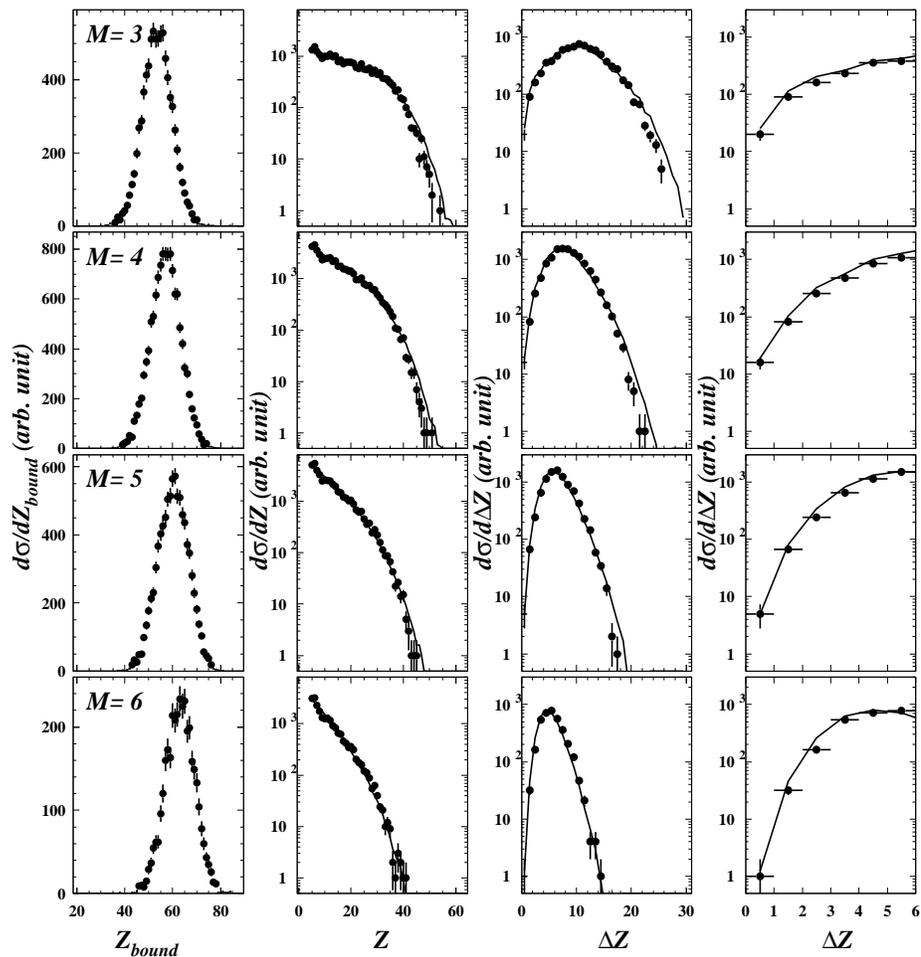,width=12.cm,height=14.5cm}
\caption {The first column shows the $Z_{bound}$ distributions for 
Xe+Sn central collision data at 32 MeV/n
for each fragment muliplicity $M=$ 3 to 6 (rows).
The following two columns superimpose $Z$ and $\Delta Z$ distributions  
of the experimental data (black symbols) and the
random break-up simulation (lines).
The last column is similar to the third one with a various range: $0 < \Delta Z < 6$.
Vertical error bars on the data points are statistical. 
The calculated distributions, performed with a 
statistics 20 times larger than the data, have been normalised. 
Error bars for the simulated distributions are small and have not been plotted. 
}
\label{fig1}
\end{center}
\end{figure}

\begin{figure}
\begin{center}
\psfig{file=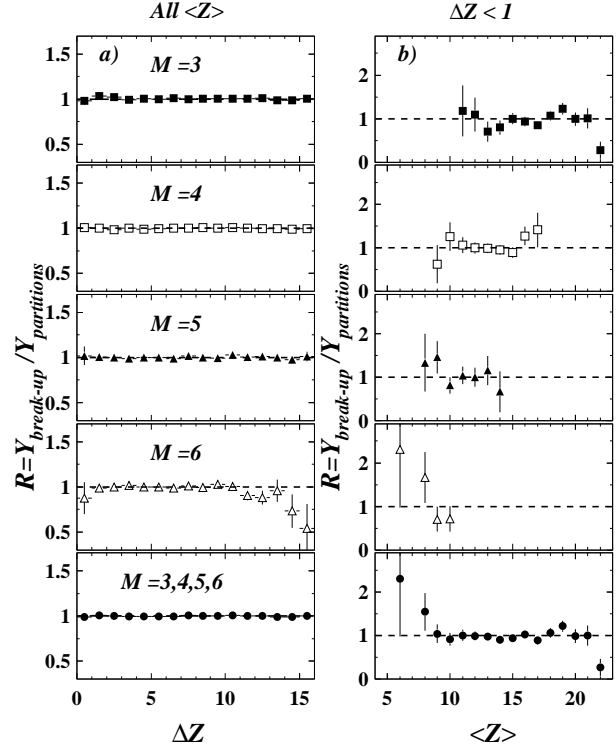,width=8cm,height=12.cm}
\caption {Ratio a) as a function of $\Delta Z$ between  the random break-up 
simulation yield and the calculation of all fragment partitions, both constrained by 
the 32 MeV/n Xe+Sn experimental $Z_{bound}$ distributions, for each multiplicity 
$M=$ 3 to 6  and their sum (rows); b) similar to a) but as a function
of $<Z>$  for the first bin in $\Delta Z$. Error bars are statistical.
}
\label{fig2}
\end{center}
\end{figure}

\begin{figure} 
\begin{center}
\psfig{file=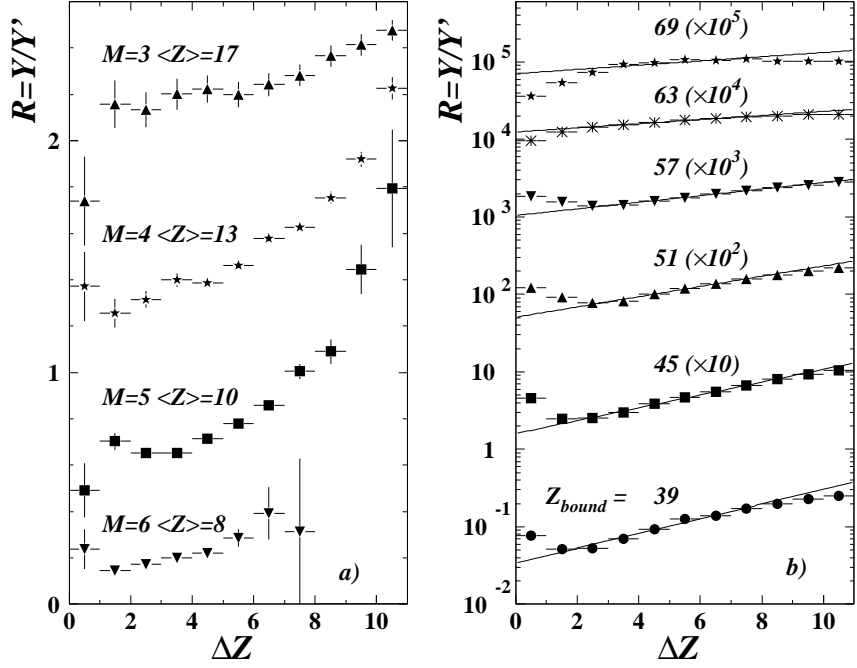,width=12.cm,height=9.6cm}
\caption {Charge correlation analysis with random break-up calculations: 
a) evolution of $R(\Delta Z,<Z>)$ as a function of $\Delta Z$ for four
($M$,$<Z>$) pairs; b) evolution of $R(\Delta Z,Z_{bound})$ as a function of $\Delta Z$
for six $Z_{bound}$ values, each one containing six atomic numbers
(scale factors are indicated in parentheses to avoid superimposition).
The straight lines  result from exponential fits in the range  $2< \Delta Z <8$. 
Vertical error bars are statistical.}
\label{fig3}
\end{center}
\end{figure}

\begin{figure} 
\begin{center}
\psfig{file=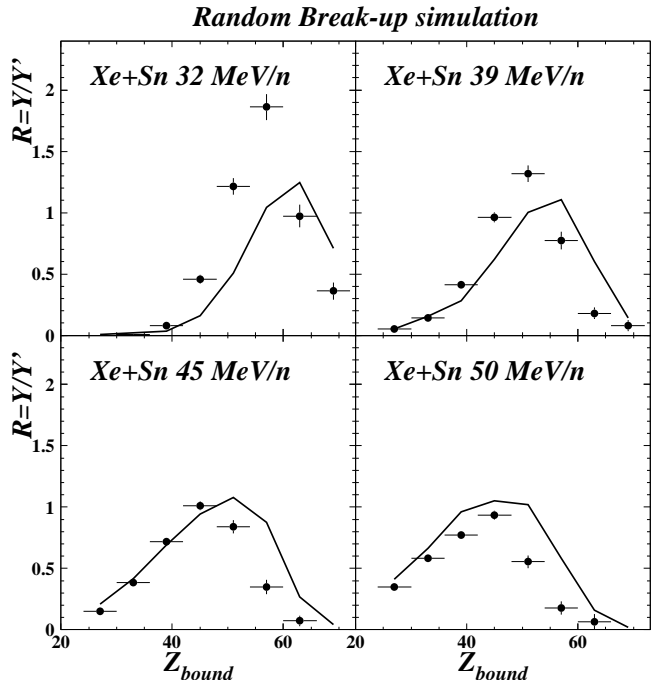,width=10.cm,height=10.cm}
\caption {Fragment charge correlations $R$ at $\Delta Z<1$ as a function of $Z_{bound}$
with random break-up calculations and ``background'' evaluations 
of \cite{borderie} (broken line).
The calculations are derived from the Xe+Sn $Z_{bound}$ distributions 
at 32, 39, 45 and 50 MeV/n.
Vertical error bars are statistical.
Horizontal bars indicate the width of the $Z_{bound}$ bin.}
\label{fig4}
\end{center}
\end{figure}

\begin{figure}
\begin{center}
\psfig{file=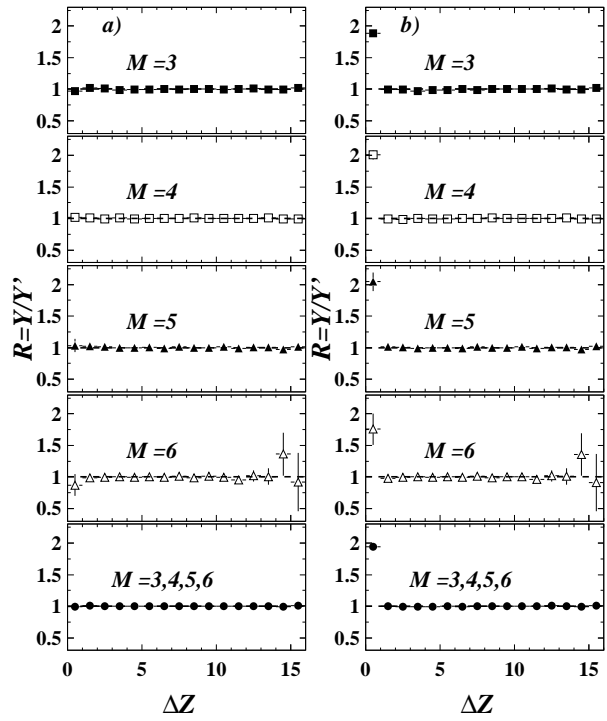,width=8.cm,height=12.cm}
\caption {Evolution of the charge correlations $R=Y/Y'$ by using the 
exchange mixing method
as a function of $\Delta Z$ with the random break-up calculations for 
a) each multiplicity from $M=3$ to $M=6$ and $M=3,4,5,6$;
b) same as a) with twice the number of events in the bin $\Delta Z<1$. 
Error bars are statistical.}
\label{fig5}
\end{center}
\end{figure}

\begin{figure}
\begin{center}
\psfig{file=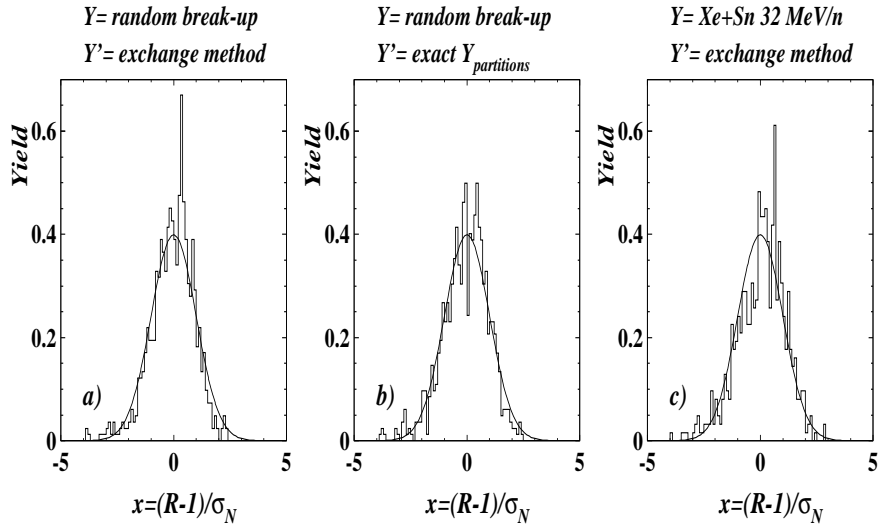,width=12.cm,height=8.cm}
\caption {Deviations from 1 of charge correlation functions $R(\Delta Z, <Z>)=Y/Y'$, 
with multiplicities $M=3,4,5$ and 6, 
divided by the statistical error $\sigma_{N}$ (numerator only).
The numerator $Y$ corresponds a,b) to the random break-up simulation; c) to the 32 MeV/n Xe+Sn data.
The denominator $Y'$ is determined by a,c) the exchange method; b) the exact
calculation of partitions.
Normal distribution (0,1) has been superimposed on every histogram.
}
\label{fig6}
\end{center}
\end{figure}

\begin{figure}
\begin{center}
\psfig{file=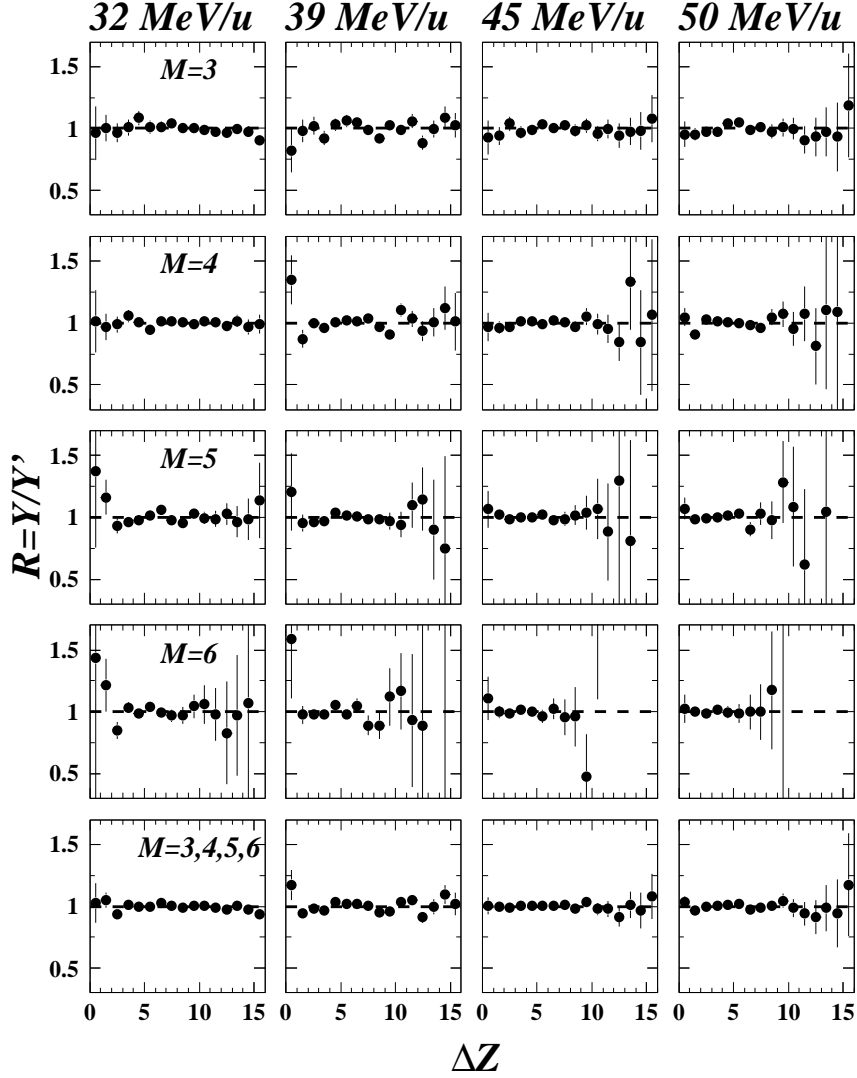,width=12cm,height=14.5cm}
\caption {Experimental charge correlations $R$ versus $\Delta Z$ for Xe+Sn central collisions,
from left to right as a function of bombarding energy (32, 39, 45 and 50 MeV/n) and
from top to bottom as a function of the fragment multiplicity, $M=3$ to 6.
The bottom row corresponds to the sum over all multiplicities.
Dotted lines indicate $R=1$.
Error bars are statistical.}
\label{fig7}
\end{center}
\end{figure}

\begin{figure}
\begin{center}
\psfig{file=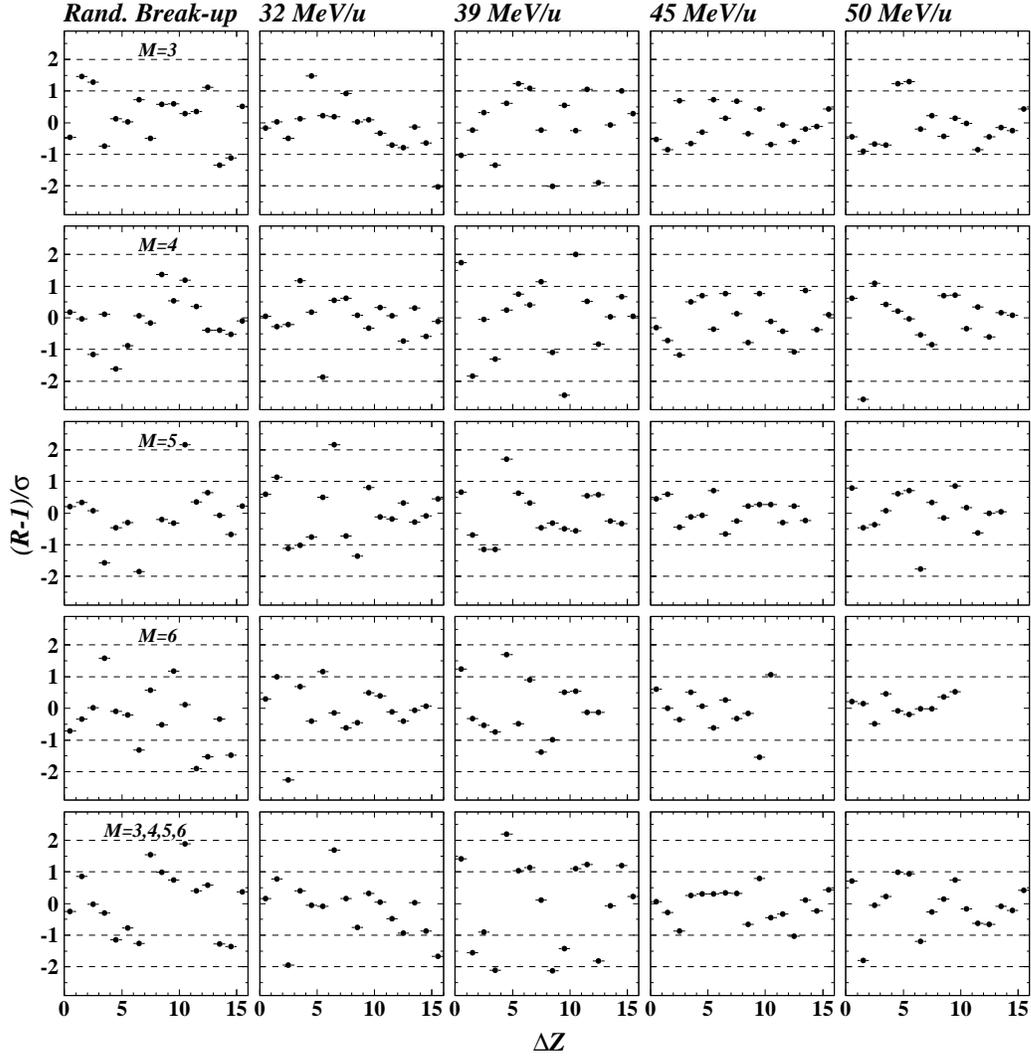,width=14.cm,height=14.cm}
\caption {Deviations from 0 of the quantity $(R-1)/\sigma$ as a function of $\Delta Z$
for simulated break-up events (first column) and experimental Xe+Sn data from 32 to 50~MeV/n
bombarding energies (columns 2 to 5), for multiplicities $M=3$ to 6 (rows 1 to 4). The last row 
corresponds to the sum over all multiplicities.
The dotted lines indicate $1\sigma$ and $2\sigma$ deviations.
}  
\label{fig8}
\end{center}
\end{figure}

\begin{figure}
\begin{center}
\psfig{file=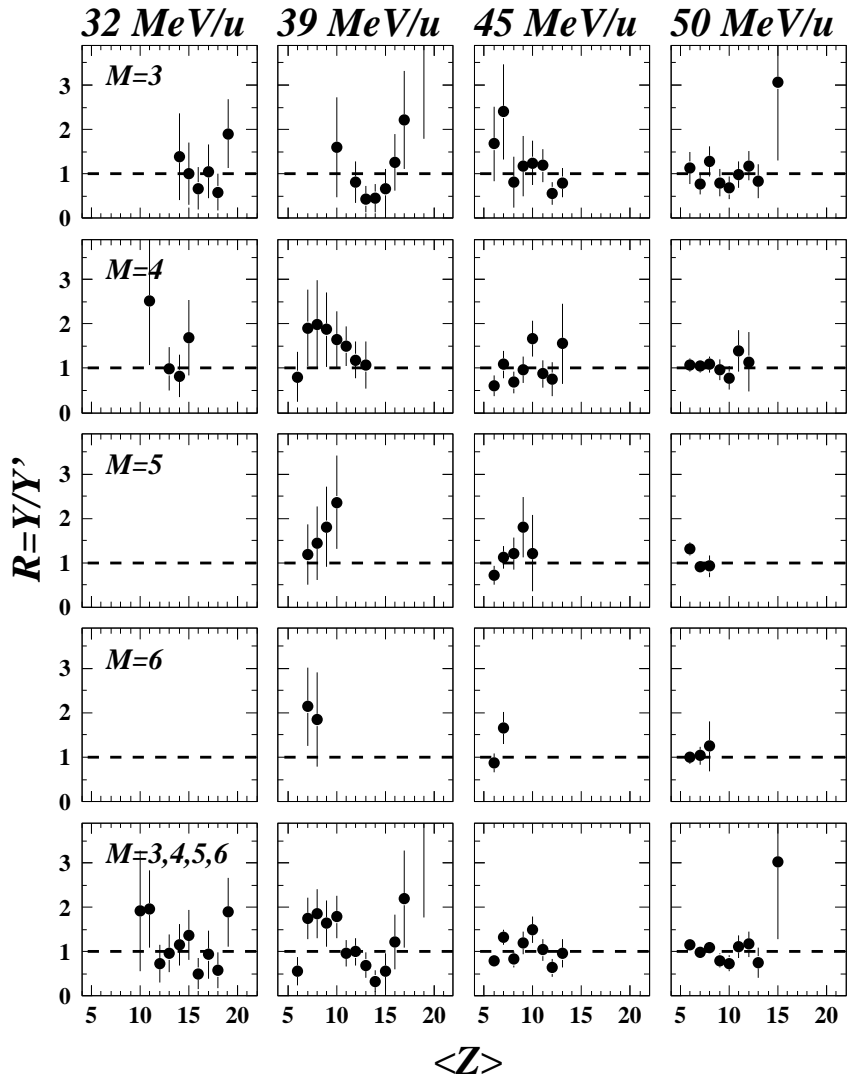,width=12.cm,height=14.5cm}
\caption {Experimental charge correlations $R$ in the bin $\Delta Z<1$ for Xe+Sn central collisions,
from left to right as a function of bombarding energy (32, 39, 45 and 50 MeV/n) and
from top to bottom as a function of the fragment multiplicity, $M=3$ to 6.
The bottom row corresponds to the sum over all multiplicities.
All data points with more than one event have been plotted.
Dotted lines indicate $R=1$.
Error bars are statistical.}
\label{fig9}
\end{center}
\end{figure}

\clearpage

\end{document}